\documentclass[lettersize,journal]{IEEEtran}

\usepackage{amsmath,amsfonts,amssymb}
\usepackage{graphicx,xcolor}
\usepackage{cite}
\usepackage{textcomp}
\usepackage{subcaption}
\usepackage{flushend}

\begin{document}

\title{Characterizing the Fault Response of the Intel Neural Compute Stick 2 Under Single-Pulse Electromagnetic Fault Injection}

\author{
Štefan Kučerák,
\thanks{Š. Kučerák and X. Hou are with the Faculty of Informatics and Information Technologies, Slovak University of Technology, Slovakia.
X. Hou is also affiliated with the State Key Laboratory of Blockchain and Data Security , Zhejiang University.
E-mail: xkucerak@stuba.sk, houxiaolu.email@gmail.com.}
Jakub Breier,
\thanks{J. Breier is with TTControl GmbH, Vienna, Austria.
E-mail: jbreier@jbreier.com.}
Xiaolu Hou}

\maketitle

\begin{abstract}
Vision processing units and other commercial neural-network inference accelerators
are increasingly deployed in safety-relevant edge applications, but their fault
response under transient hardware disturbances remains poorly characterized in
the open literature. For the Intel Movidius Myriad~X, packaged as the Intel
Neural Compute Stick~2 (NCS2), only a single feasibility study has been
published. We report a systematic single-pulse electromagnetic fault injection
(EMFI) campaign on the NCS2 running three ImageNet-trained convolutional neural
networks (ResNet-18, ResNet-50, VGG-11) on the OpenVINO runtime. Across 1{,}536 spot-test trials at characterized hotspots and approximately 16{,}000 parameter-search trials, single pulses produce four reproducible outcome classes:
no measured accuracy change, minor silent data corruption, major persistent degradation that survives across subsequent inferences until model reload, and device hangs requiring USB power-cycling; these outcomes are respectively interpreted as no-effect, SDC with possible SET-like or small persistent-state mechanisms, SEU-like persistent corruption, and SEFI-like loss of functionality.
Two findings are central. First, the major-degradation class can be induced at
18--31\,\% of trials at characterized hotspots, with post-collapse top-1
accuracy below five percent and persistence across all subsequent inferences
until explicit model reload---a regime that no inference-API-level mechanism
detects. Second, this regime is also inducible by pulses delivered to an idle
device with the model already loaded, demonstrating that load-time integrity
checks alone are insufficient. The three evaluated architectures occupy
distinguishable regions of the persistent-fault-rate--device-failure-rate plane,
establishing model-architecture choice as a deployment-time reliability lever.
We discuss mitigation strategies graded by class, focusing on mechanisms implementable at the application level without modification to the device firmware or the OpenVINO runtime.
\end{abstract}


\section{Introduction}
\label{sec:intro}

\IEEEPARstart{D}{eep}-learning inference is moving from datacenter accelerators to small, low-power, fixed-function devices that sit close to the sensor: vision processing units (VPUs), neural-network coprocessors, and dedicated inference sticks. These devices are no longer confined to laboratory prototyping. They appear in industrial inspection lines, in research vehicles, in mobile robotics platforms, and increasingly in deployed edge perception pipelines that feed downstream control systems. In several of these deployment domains the inference output participates, directly or indirectly, in safety-relevant decisions, and the device that produced it falls under the scope of functional-safety standards such as IEC~61508~\cite{iec61508} and ISO~26262~\cite{iso26262}, with additional guidance for AI/ML components emerging through ISO/IEC~TR~5469~\cite{isoiec5469} and ISO~21448~\cite{iso21448}. These standards do not require zero failure. They require that failure modes be \emph{characterized}, that residual fault rates be quantified to a level consistent with the intended safety integrity, and that the system architecture provide detection or mitigation for failure modes that exceed the budget.

Commercial inference accelerators present a problem for this kind of argument. Their internal architectures are partially or fully proprietary, their failure characteristics under transient hardware disturbances are not generally published by the vendor, and most of the published reliability literature on them concerns datacenter GPUs or FPGA-based accelerators rather than the small fixed-function silicon found at the edge. For the Intel Movidius Myriad~X---packaged as the Intel Neural Compute Stick~2 (NCS2) and used as an off-the-shelf development and prototyping target across the deployment domains listed above---the published characterization is limited to a single late-breaking-results paper~\cite{bhasin2025practical}, which establishes that electromagnetic disturbances can produce inference errors but does not characterize the resulting fault space.

The present work fills this gap. We use single-pulse electromagnetic fault injection (EMFI) as a controlled laboratory disturbance source and report a systematic characterization of the NCS2 running three production-scale convolutional neural networks on the OpenVINO runtime. EMFI is a near-field disturbance technique in which a small magnetic coil placed close to a packaged integrated circuit delivers a brief, localized pulse that induces transient currents in on-die wiring~\cite{oflynn2021picking}. The induced disturbance can flip individual bits in registers and embedded memory, briefly violate combinational-logic timing, or perturb control state, producing fault patterns that overlap qualitatively with single-event upsets, single-event transients, and single-event functional interrupts as classified in the radiation-effects literature~\cite{oflynn2021picking}. Compared with neutron-beam testing, EMFI is several orders of magnitude cheaper per pulse and allows the position of the disturbance to be controlled, which together make it a practical tool for characterization studies that would be infeasible at a beamline.

\subsection{Contributions}

This paper makes the following contributions:

\begin{enumerate}
\item A four-class fault outcome taxonomy for single-pulse EMFI on the NCS2---unchanged inference, minor accuracy degradation (SDC), major persistent degradation, and device hangs---with phenomenology-grounded mapping to the SET/SEU/SEFI vocabulary of the radiation-effects literature.

\item A spatial fault map across the package surface using three near-field probe geometries, identifying two reproducible lateral hotspots separated by approximately 7~mm and a central region that preferentially produces device hangs.

\item Evidence that loaded model state on the NCS2 is fault-susceptible while the device is otherwise idle: pulses delivered \emph{before} inference produce persistent misclassification at rates comparable to during-inference pulses, implying that load-time integrity checks alone are insufficient.

\item A direct comparison across three CNN architectures (ResNet-18, ResNet-50, VGG-11) under identical pulse parameters, showing that architecture choice systematically shifts the device into different regions of the persistent-fault-rate--device-failure-rate plane.

\item Day-to-day repeatability of the outcome distribution at parameter points inside the fault-onset region, addressing a standard threat to validity in EMFI characterization studies.

\item A graded set of mitigation strategies derived from the taxonomy (host-side liveness watchdog, reference-image cross-checking, redundant-inference voting), implementable at the application level without modification to the device firmware or the OpenVINO runtime.
\end{enumerate}

\subsection{Paper Organization}

Section~\ref{sec:background} reviews the NCS2 platform and surveys related work on hardware-induced faults in neural-network accelerators. Section~\ref{sec:setup} describes the experimental setup, the target preparation, and the EMFI platform. Section~\ref{sec:method} gives the methodology, the fault taxonomy, and the evaluation metrics. Section~\ref{sec:results} reports the experimental results. Section~\ref{sec:discussion} discusses reliability implications, mitigation strategies, and security implications. Section~\ref{sec:threats} identifies threats to the validity of the results. Section~\ref{sec:concl} concludes and identifies follow-up work.

\section{Background and Related Work}\label{sec:background}

\subsection{The Intel Neural Compute Stick 2 and the Myriad X VPU}

The Intel Neural Compute Stick~2~\cite{intel_ncs2_brief} is a USB~3.0 form-factor inference accelerator built around the Intel Movidius Myriad~X system-on-chip; the Myriad~X is the third-generation member of a VPU family whose architectural blueprint was introduced with Myriad~2~\cite{moloney2014myriad}. The Myriad~X integrates a small number of LEON RISC cores responsible for control flow and host communication, sixteen vector processors known as SHAVEs (Streaming Hybrid Architecture Vector Engine) responsible for the bulk of arithmetic in convolutional layers, dedicated neural compute and imaging engines, and on-die CMX (connection matrix) memory that holds activations and partial results during inference. Models compiled by the OpenVINO toolchain are transferred over USB into device memory, after which the host issues inference requests through the OpenVINO runtime API and receives back numerical outputs. The device exposes no published debug interface for reading back internal state, and the floorplan of the Myriad~X die is not publicly available.

The NCS2 is cooled passively. Sustained inference loads cause the package surface to warm appreciably, and we observed during preliminary characterization that thermal throttling, when allowed to occur, modulated the operating frequency and produced noticeable variability in fault rates between trials. The experimental setup described in Section~\ref{sec:setup} therefore uses an external fan to suppress throttling during long campaigns.

The package geometry of the NCS2 has implications for the EMFI campaign that are worth stating in advance. The Myriad~X die is approximately $8\times 8$~mm; the 4~mm CCW probe used in Section~\ref{sec:results} therefore couples energy into a substantial fraction of the die surface. A 1~mm probe is the appropriate spatial-resolution choice for characterization, while the 4~mm probe is useful primarily for understanding the device response when a substantial fraction of the die is disturbed simultaneously. We discuss this asymmetry quantitatively in Section~\ref{sec:repeatability}.

\subsection{Electromagnetic Fault Injection as a Reliability Surrogate}

EMFI is a near-field magnetic-coupling technique in which a small coil placed in close proximity to a packaged integrated circuit delivers a short high-voltage pulse, inducing transient currents in on-die wiring that propagate into local logic and storage. Industrial-grade injectors such as NewAE's ChipSHOUTER and Riscure's EMFI tool are in widespread laboratory use~\cite{oflynn2021picking}. The detailed physical mechanism by which an EMFI pulse becomes a logic-level fault depends on the specific package, supply network, and pulse parameters, and is not in general fully understood for any commercial target~\cite{oflynn2021picking}.

EMFI provides a controlled, repeatable, low-cost laboratory surrogate for the broader class of transient hardware disturbances that commercial accelerators may experience in deployment, including single-event upsets and transients due to terrestrial neutron flux, supply-voltage glitches, and electromagnetic interference. The phenomenological mapping between EMFI-induced faults and field-relevant transient disturbances has been examined explicitly by O'Flynn~\cite{oflynn2021picking}, who argues that the fault classes produced by EMFI overlap substantially with those produced by alternative perturbation sources. 
Hou~\textit{et~al.}~\cite{hou2021physical} provide a complementary comparative evaluation across multiple physical fault-injection vectors specifically on edge deep-learning hardware, finding that different injection channels can drive overlapping but distinct failure distributions.
EMFI is not, however, a substitute for a neutron-beam study where one is feasible: the present results should be interpreted as a fault-response characterization under EMFI rather than as a quantitative prediction of the soft-error rate observed in deployment. The mitigation strategies discussed in Section~\ref{sec:discussion} depend only on the four fault classes identified in Section~\ref{sec:fault-classification}, which we expect to be qualitatively preserved across different transient-disturbance sources.

\subsection{Fault Injection on Neural-Network Accelerators}

Hardware-induced faults in neural-network inference have been studied at three increasingly fine-grained levels: simulation, FPGA prototype, and packaged silicon.

At the simulation level, studies have characterized the response of trained convolutional networks to random bit flips in weights and activations. Li~\textit{et~al.}~\cite{li2017understanding} examined error propagation through convolutional architectures and identified a strong dependence on the bit position of the flipped bit, with high-order exponent positions in floating-point weights producing disproportionately large output deviations. Reagen~\textit{et~al.}~\cite{reagen2018ares} extended this to a broader range of architectures and produced quantitative resilience metrics. Mahmoud~\textit{et~al.}~\cite{mahmoud2020pytorchfi} packaged similar functionality as a reusable PyTorch-level fault-injection tool. Hong~\textit{et~al.}~\cite{hong2019terminal} demonstrated that single bit flips in the most significant bits of weight tensors can drive a network to near-random output. These simulation results provide a useful baseline against which hardware-induced faults can be compared, but cannot reproduce control-flow corruption (a bit flip in a SHAVE program counter or DMA descriptor) or the interaction of a transient disturbance with whichever tensor happens to be in transit at a particular moment.

At the FPGA-prototype level, Khoshavi~\textit{et~al.}~\cite{khoshavi2020fpgann} examined the resilience of DNN inference under voltage underscaling on commercial FPGAs and identified layer-dependent sensitivity profiles. The FPGA-level injection channels (configuration-memory bit flips, supply undervolting, partial reconfiguration) differ from those available on packaged ASIC accelerators such as the Myriad~X, and the resulting fault rates do not directly translate.

At the packaged-silicon level, fault-injection studies on commercial neural-inference hardware are sparse. Breier~\textit{et~al.}~\cite{breier2018practical} demonstrated practical fault induction on a microcontroller-class deep-learning implementation using laser injection, and subsequently showed that sign-bit-flip faults can be exploited to reverse-engineer the parameters of a trained network~\cite{breier2021sniff}.
Liu~\textit{et~al.}~\cite{liu2017fault} examined the response of a deep neural network to fault injection at the architecture level. 
Breier~\textit{et~al.}~\cite{breier2026weight} investigated the influence of model parameter number representations to fault resistance by using EMFI.
The closest prior work to the present study is the two-page late-breaking-results paper by Bhasin~\textit{et~al.}~\cite{bhasin2025practical}, which establishes the feasibility of inducing inference errors on the NCS2 with multiple consecutive pulses on a small reference network, but does not characterize spatial sensitivity, fault persistence beyond the perturbed inference, architectural dependence across CNN families, or methodological confounds such as thermal state and day-to-day variation.

\section{Experimental Setup}\label{sec:setup}
\begin{figure}[!t]
    \centering
    \includegraphics[width=0.98\linewidth]{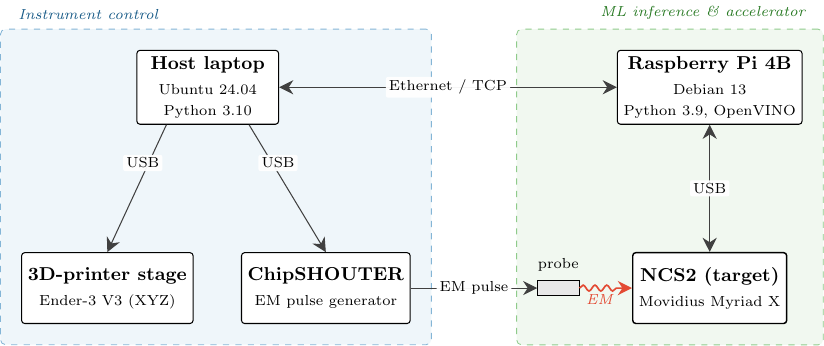}
    \caption{Experimental setup overview. The host laptop (left) controls
  instrument hardware---the ChipSHOUTER and the XYZ stage---while the
  Raspberry~Pi (right) hosts the OpenVINO inference stack and
  communicates with the NCS2 target over USB~3.0. The two control hosts
  exchange commands over a dedicated Ethernet link. The ChipSHOUTER drives a near-field probe positioned by the stage above
  the exposed Myriad~X package on the decased NCS2 board; the EM pulse couples into the package surface
  (red wavy arrow) without electrical contact.}
    \label{fig:setup}
\end{figure}

The experimental setup schematic is depicted in Fig.~\ref{fig:setup}.
This section provides details on each of the setup components.

\subsection{Target Platform}
\label{subsec:target_platform}

The target platform of this study is the Intel Neural Compute Stick 2 (NCS2), a USB-connected neural-network inference accelerator. The NCS2 was selected as a representative edge-AI accelerator because it supports low-power execution of convolutional neural networks while providing limited visibility into the internal execution state of the accelerator. This makes it a suitable platform for evaluating the reliability of neural-network inference under externally induced physical faults.

The experimental control architecture consisted of an Ubuntu-based host laptop, a Raspberry~Pi~4B with 4~GB of RAM, and the Intel NCS2. The host laptop (AMD Ryzen~7~5800H, 16~GB RAM, Ubuntu~24.04~LTS) acted as the main experiment controller, coordinating the experiment flow and collecting logs. The Raspberry~Pi~4B (Debian~GNU/Linux~13) was connected to the host through a dedicated Ethernet link and executed the inference workload on the NCS2 over USB~3.0; OpenVINO~build~2022.3.2 for ARM64~Linux drove inference through the \texttt{MYRIAD} device target. The Pi was inserted into the control architecture specifically to enable software-controlled USB power cycling of the NCS2, which was necessary because some injected faults caused the accelerator to disconnect or enter unrecoverable states.

The software stack was implemented primarily in Python using a client--server architecture. The host laptop operated as the client, while the Raspberry Pi executed a server process that controlled inference on the NCS2. Communication between the two devices was implemented over TCP sockets. For each experiment, the host sent a command to the Raspberry Pi to start the inference run; after completion, the Pi returned the measured top-1 and top-5 accuracy values, providing a coarse software-level synchronization point between the experiment controller and the inference process.

The Python environment was deliberately partitioned across the two control hosts to keep all neural-network and accelerator code on a single machine. The Raspberry~Pi hosted the full inference stack (OpenVINO together with PyTorch, TorchVision, TorchMetrics, and NumPy), while the host laptop ran only stage-motion control over the 3D-printer G-code interface, ChipSHOUTER trigger and parameter control over its USB serial interface, and TCP-socket exchange of run commands and accuracy with the Raspberry~Pi; no inference, model code, or NCS2-facing call paths existed on the host laptop. The campaign also depended on Optuna~4.7.0 for parameter-search trial selection. Exact library versions are listed in the supporting repository. We attempted to obtain the NCS2 firmware/runtime version through the OpenVINO and \texttt{usbutils} interfaces but did not find a documented way to query it; the NCS2 was used in its as-shipped configuration without firmware modification.

To improve the robustness of long-running experiments, a watchdog mechanism on the Raspberry~Pi recovered the device by power-cycling its USB port via the \texttt{uhubctl} utility whenever the NCS2 did not respond within 5~s. This automated recovery from many fault-induced hangs without manual reconnection. After each model reload, a reference inference was performed and the output compared with a stored golden output to verify correct loading before the next experiment.

\begin{figure}
    \centering
    \includegraphics[width=0.98\linewidth]{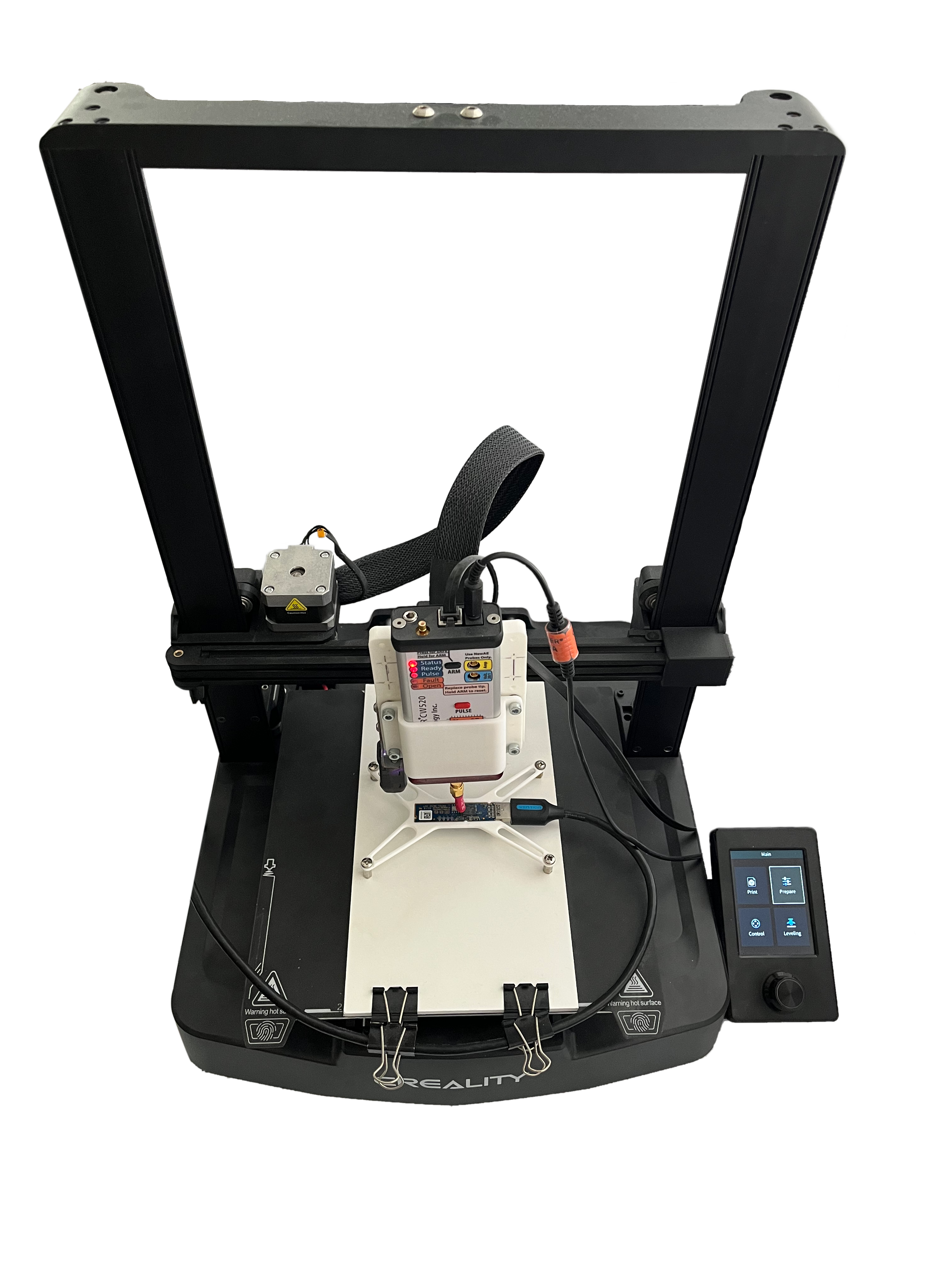}
    \caption{Decased Intel Neural Compute Stick~2 placed on the evaluation platform, with the Myriad~X package accessible for probe positioning.}
    \label{fig:ncs2}
\end{figure}

The NCS2 was physically prepared by manually removing the device casing to expose the printed circuit board and passive heatsink, enabling repeatable positioning of the electromagnetic probe above the target area. The exposed device was mounted on a custom PLA holder fabricated to match the experimental bed mounting holes, providing mechanical stability and improving probe-positioning repeatability. Active cooling, provided by an ADDA AD0812HS-A70GL 12~V DC brushless fan placed approximately 5~cm laterally from the target SoC and operated at constant 12~V, was used to suppress thermal throttling during long campaigns. The ambient laboratory temperature was not directly logged; we discuss the implications in Section~\ref{sec:threats}. When available, the OpenVINO interface was used to read the NCS2 thermal information through the \texttt{DEVICE\_THERMAL} property of the \texttt{MYRIAD} device. The NCS2 within the evaluation platform is depicted in Fig.~\ref{fig:ncs2}.

\subsection{Electromagnetic Injection Platform}
\label{subsec:emfi_platform}

The electromagnetic fault injection platform was built around the NewAE ChipSHOUTER pulse generator, which produces short high-voltage pulses that drive a near-field magnetic probe placed in close proximity to the target package. The pulse generator was controlled by the host PC over a USB serial link using the Python \texttt{chipshouter} library; the pulse parameters relevant to the present single-pulse study are the charge voltage and the pulse width. Three interchangeable near-field probes were used: a 1~mm clockwise (CW) coil, a 1~mm counter-clockwise (CCW) coil, and a 4~mm CCW coil. The two 1~mm probes provide finer lateral localization than the 4~mm CCW probe, while the 4~mm CCW probe couples energy over a larger area and was used to investigate the response of the device to broader-area disturbances.

Probe positioning was controlled by a three-axis stage derived from a Prusa~MK2 3D-printer mechanism. The stage was driven over USB serial at 115200~baud using a standard G-code interface; absolute positioning commands (\texttt{G90}/\texttt{G0}) were issued together with a synchronous wait-for-completion (\texttt{M400}) to ensure that subsequent pulse delivery occurred only after the stage had reached the commanded coordinates. The lateral $(x,y)$ coordinate system was defined with respect to the bench frame and was kept consistent across all experiments, so that absolute coordinates are comparable between campaigns reported in this paper. The stand-off distance $z$ between the probe tip and the package surface depends on the probe used because of differences in tip geometry; probe-specific $z$-offsets, calibrated in advance, are listed in Table~\ref{tab:probe-zoffsets}. All $z$ values reported in this paper refer to the corrected probe-tip-to-package distance after subtraction of the probe-specific offset; values reported in the supporting repository at the time of the original campaign use the uncorrected stage-encoder scale and are related to the corrected scale by an additional 0.45~mm offset for the experiments performed before the calibration was finalized.

\begin{table}[!t]
\caption{Probe-specific $z$-offset between the stage encoder origin and the
probe-tip-to-package datum. Negative values indicate that the probe tip
extends beyond the encoder origin in the negative-$z$ direction.}
\label{tab:probe-zoffsets}
\centering
\begin{tabular}{lc}
\hline
Probe & $z$-offset (mm) \\
\hline
1~mm CCW & $-1.05$ \\
1~mm CW  & $-0.85$ \\
4~mm CCW & $-1.75$ \\
\hline
\end{tabular}
\end{table}

The synchronization between the EM pulse and the inference workload was implemented in software. The host PC orchestrated each trial by exchanging short text messages with the Raspberry~Pi over a TCP socket on a dedicated Ethernet link, and the ChipSHOUTER trigger was issued by the host PC over the USB serial link using the \texttt{chipshouter} library. The pulse-trigger command and the inference-start command are therefore separated by an OS-mediated control path that includes Python interpreter scheduling, USB-stack latency, and TCP round-trip time. The combined latency is on the order of milliseconds and is not directly measured. The configuration is therefore not suitable for resolving fault sensitivity at the granularity of individual neural-network layers, and the present study does not attempt such resolution. A hardware-triggered configuration, in which a GPIO line asserted by an instrumented OpenVINO build directly clocks the ChipSHOUTER input, is identified as a primary methodological improvement for follow-up work. To partially compensate for the trigger jitter, every reported result corresponds to a fixed nominal delay relative to the inference start (1~s before or 1~s after, see Section~\ref{sec:campaign}), and within each campaign the delay was held constant.

Two scanning regimes were used. In the first regime, parameter ranges were defined for the lateral coordinates $(x, y) \in [113, 127] \times [148, 160]$~mm, the stand-off $z \in [0, 2]$~mm, and the pulse charge voltage $V \in [150, 500]$~V; trial points within this space were selected by a Tree-structured Parzen Estimator (TPE) sampler, as described in Section~\ref{sec:campaign}. In the second regime, used for repeatability and architecture-comparison experiments, a single $(x, y, z, V)$ configuration was held constant across many trials and only the inference workload was varied. Unless otherwise stated, all reported quantitative results in Section~\ref{sec:results} correspond to a single pulse per trial with pulse width 160~ns; repeated-pulse experiments were performed during early exploration and are not used in the quantitative analysis.

\subsection{Neural-Network Workloads}
\label{subsec:nn_workloads}

The experimental campaign evaluated image-classification workloads executed on the Intel Neural Compute Stick 2. Three convolutional neural-network architectures were considered: ResNet-18, ResNet-50, and VGG-11. These models were selected to include networks with different depths and architectural structures. ResNet-18 and ResNet-50 represent residual-network architectures with different numbers of layers, whereas VGG-11 follows a more sequential convolutional structure with fully connected layers at the end of the network. This selection makes it possible to compare EMFI-induced behavior across different model architectures rather than limiting the evaluation to a single neural-network design.

All evaluated models were trained for ImageNet classification and were executed using the OpenVINO inference pipeline. The experiments used images from the ImageNet validation dataset, preprocessed according to the input requirements of each deployed model. Each input image was classified independently, and the predicted output was compared with the expected label. The primary evaluation metrics were top-1 and top-5 accuracy.

Different image-subset sizes were used depending on the purpose of the experiment. Smaller subsets (64, 128, and 256 validation images) were used during exploratory searches to reduce experiment time and allow more parameter combinations to be tested. The main spot-testing campaign used 512 ImageNet validation images per evaluated model, with each main configuration repeated for 256 trials.

Two inference modes were considered. In the standard \emph{synchronous} mode, images were processed one at a time, and the corresponding classification result was collected before continuing with the next image. This mode was used for the main evaluation because it provides a direct relationship between the injected fault and the observed classification outcome. \emph{Asynchronous} inference, in which multiple inference requests are queued and executed in overlap, was evaluated for selected models because it more closely reflects higher-throughput deployment scenarios and may expose different fault behavior due to overlapping computation, buffering, or scheduling inside the inference pipeline.

\section{Methodology}\label{sec:method}

\subsection{Baseline Measurement}

A baseline characterization of the unperturbed system was performed for each evaluated model before any fault-injection campaign was started. The baseline procedure consisted of loading the model into the NCS2, performing a reference inference on a fixed image to verify that the model had been loaded correctly, and then evaluating top-1 and top-5 accuracy over the selected validation subset under nominal conditions, with no electromagnetic pulse delivered. The validation subset comprised 512 randomly drawn images from the ImageNet validation set, with the random subset fixed by a deterministic seed (seed value 21 in the random-split used by \texttt{torchvision}); the same subset was reused for the perturbed campaigns, so that any difference in measured accuracy can be attributed to the EM pulse rather than to dataset-level variation. Baseline measurements were repeated at the start of each experiment day to detect drift in the unperturbed behavior of the device; no significant drift was observed across days, although day-to-day variability of the perturbed behavior is examined in Section~\ref{sec:repeatability}. The baseline accuracies obtained for each model are reported in Section~\ref{sec:baseline-results} and serve as the reference against which all subsequent perturbed measurements are compared.

\subsection{EMFI Campaign Procedure}
\label{sec:campaign}

The fault-injection campaign was organized in two phases. The first phase was an exploratory scan of the parameter space, intended to identify spatial and parametric regions in which the EM pulse produced an effect on the inference output. The second phase consisted of \emph{spot tests} at parameter configurations identified in the first phase, in which the pulse parameters were held constant and the inference workload was repeated many times to obtain a stable estimate of the per-trial outcome distribution. This separation between exploration and characterization is important: the exploratory phase is a search procedure biased toward producing faults and therefore does not provide an unbiased estimate of fault rates, while the spot-test phase produces such estimates at the cost of restricting attention to a small number of parameter points.

A single trial within either phase consisted of the following sequence. The selected model was loaded onto the NCS2 over USB. A reference image was then classified, and the predicted label was compared against a stored golden output to verify that the model had been loaded correctly; if the verification failed, the trial was discarded and the device was power-cycled by software through the Raspberry~Pi. The probe was then moved to the configured $(x, y, z)$ position and the ChipSHOUTER was charged to the configured voltage. The inference workload was then started and a single EM pulse was delivered at the configured timing relative to the start of inference. The pulse-versus-inference timing is controlled in software by a signed scalar \emph{delay}~$d$: for $d > 0$, the inference is started first and the pulse is delivered $d$~seconds later; for $d < 0$, the pulse is delivered first and the inference is started $|d|$~seconds afterward. All quantitative results in Section~\ref{sec:results} correspond to $d = +1$~s (during-inference) or $d = -1$~s (before-inference). The inference results, the device-status flags returned by the OpenVINO runtime, and the elapsed time were logged together with the full trial configuration to a CSV record. Between trials, the model was explicitly reloaded so that any persistent corruption of the loaded model state did not contaminate subsequent measurements.

For the exploratory phase, the parameter space $(x, y, z, V, d)$ is too large to be exhausted by grid search: a 1~mm lateral grid combined with a 0.05~mm $z$-resolution and a 10~V voltage step over the ranges given in Section~\ref{subsec:emfi_platform} already yields more than $10^{6}$ candidate points, of which only a small minority produce a measurable effect. The Optuna~\cite{akiba2019optuna} hyperparameter optimization framework was therefore used to drive trial selection. The TPE sampler was configured with multivariate group sampling, allowing correlations between parameters to be exploited during search. The objective passed to the sampler was the absolute deviation of measured top-1 accuracy from the baseline accuracy of the same model on the same image subset, so that both accuracy drops and any unexpected accuracy increases would be reported as informative trials. The sampler was run separately for each combination of model, probe, and inference mode under investigation.

The spot tests were performed at parameter configurations selected from the Optuna runs on the basis of two criteria: a high observed rate of trials with non-zero accuracy deviation, and a low observed rate of trials in which the device entered a state requiring USB power-cycling. Two such configurations were identified for the 1~mm CCW probe and used for the model comparison reported in Section~\ref{sec:arch-results}; the spot-test configuration used for the principal model-comparison campaign is the right-side hotspot at $(x, y, z) = (123.4, 155.1, 0.25)$~mm, with charge voltage $V = 348$~V, pulse width 160~ns, and a single pulse per trial. At each spot-test configuration, the trial sequence described above was repeated 256 times per model with the inference workload set to 512 images from the validation set; the per-trial outcome of all 256 trials forms the basis for the histograms presented in Section~\ref{sec:results}.

\subsection{Fault Classification}
\label{sec:fault-classification}

The output of each fault-injection trial was assigned to one of four mutually exclusive outcome classes, defined to support the SET--SEU mapping used throughout this paper. Class~$C_{0}$, \emph{unchanged}, is assigned to a trial in which inference completes and the measured top-1 accuracy over the inference workload is within $\theta_{\mathrm{minor}}$ of the unperturbed baseline. Class~$C_{1}$, \emph{minor accuracy perturbation}, is assigned to a trial in which inference completes and the measured top-1 accuracy differs from baseline by between $\theta_{\mathrm{minor}}$ and $\theta_{\mathrm{major}}$; the deviation may be transient and confined to the perturbed inference, or it may persist into subsequent inferences within the same trial without crossing the major-degradation threshold. Class~$C_{2}$, \emph{major persistent degradation}, is assigned to a trial in which the measured top-1 accuracy collapses by more than $\theta_{\mathrm{major}}$ and remains collapsed across all subsequent inferences within the trial until the model is explicitly reloaded over USB. Class~$C_{3}$, \emph{device failure}, is assigned to a trial in which the device fails to return a valid result within the watchdog timeout and requires a USB power-cycle to recover; this includes hangs, USB disconnections, and OpenVINO runtime errors.

The thresholds $\theta_{\mathrm{minor}} = 0.01$ and $\theta_{\mathrm{major}} = 0.50$ in top-1 accuracy were chosen empirically. The lower threshold corresponds to approximately five images out of the 512-image workload and is therefore larger than the single-image accuracy quantum of $1/512$; trials within $\theta_{\mathrm{minor}}$ of the baseline are treated as $C_{0}$ for binning purposes rather than as a separately resolved fault class. The upper threshold $\theta_{\mathrm{major}} = 0.50$ was chosen on the basis of the visibly bimodal structure of the spot-test distributions reported in Section~\ref{sec:spot-results}: the mass between baseline and baseline$\,-\,0.50$ is small for the ResNet models and constitutes the genuinely intermediate $C_{1}$ regime, while trials below baseline$\,-\,0.50$ cluster sharply near top-1 accuracy of zero and form the $C_{2}$ population. We note that the $\theta_{\mathrm{minor}}$ threshold is not directly comparable across workloads of different sizes because a single per-image misclassification translates to an accuracy change of $1/N_{\mathrm{img}}$; this implication is examined when we vary $N_{\mathrm{img}}$ in Section~\ref{sec:image-count}.

The four classes admit a natural mapping to the SET and SEU vocabulary established in the radiation-effects literature. Class~$C_{1}$ is interpreted as a minor SDC outcome that may arise either from a SET-like transient affecting the in-flight computation or from a small persistent corruption of stored state whose effect remains below the major-degradation threshold. Class~$C_{2}$ is interpreted as an SEU-like outcome: a persistent state change in loaded model parameters, compiled instructions, or accelerator configuration that survives until explicit recovery. Class~$C_{3}$, in radiation-effects terminology, corresponds most closely to a single-event functional interrupt (SEFI), in which a transient disturbance triggers a persistent loss of device functionality rather than a localized data corruption. We emphasize that classes $C_{1}$ and $C_{2}$ together constitute the \emph{silent} failure population from the perspective of a host application that receives only a successful inference return and lacks an independent correctness check: the inference call returns normally and produces a result that may be slightly or severely incorrect. Class~$C_{3}$, in contrast, is detectable by any reasonable host-side liveness check.

\subsection{Evaluation Metrics}

Two distinct levels of accounting are used in the evaluation. At the level of an individual trial, classification correctness over the image workload is summarized by the top-1 and top-5 accuracy metrics:
\begin{equation}
A_{\mathrm{top1}} = \frac{N_{\mathrm{correct,top1}}}{N_{\mathrm{total}}}, \qquad
A_{\mathrm{top5}} = \frac{N_{\mathrm{correct,top5}}}{N_{\mathrm{total}}},
\label{eq:acc}
\end{equation}
where $N_{\mathrm{total}}$ is the number of inferences in the workload and $N_{\mathrm{correct,top1}}$ ($N_{\mathrm{correct,top5}}$) counts inferences whose top-1 (top-5) prediction matches the ground-truth label.

At the level of the campaign as a whole, fault outcomes are summarized by the misclassification rate, the device-failure rate, and the persistent-fault rate:
\begin{equation}
R_{\mathrm{mis}} = \frac{N_{C_1}+N_{C_2}}{N_{C_0}+N_{C_1}+N_{C_2}}, \quad
R_{\mathrm{fail}} = \frac{N_{C_3}}{N_{\mathrm{trial}}}, \quad
R_{\mathrm{persist}} = \frac{N_{C_2}}{N_{\mathrm{trial}}},
\label{eq:rates}
\end{equation}
where $N_{\mathrm{trial}}=N_{C_0}+N_{C_1}+N_{C_2}+N_{C_3}$ denotes the number of trials in the campaign.

The misclassification rate $R_{\mathrm{mis}}$ is computed only over trials that completed without device failure, since trials in class~$C_{3}$ do not produce a meaningful classification output. Reporting $R_{\mathrm{mis}}$ in isolation is therefore insufficient: a configuration that produces a high $R_{\mathrm{fail}}$ may exhibit a misleadingly low $R_{\mathrm{mis}}$ simply because most of its trials were censored. We therefore report $R_{\mathrm{mis}}$, $R_{\mathrm{fail}}$, and $R_{\mathrm{persist}}$ jointly throughout Section~\ref{sec:results}, and rely on the four-class taxonomy of Section~\ref{sec:fault-classification} as the primary description of the fault response.

\section{Experimental Results}\label{sec:results}

\subsection{Baseline Accuracy}
\label{sec:baseline-results}

The baseline characterization of the three evaluated models was performed under nominal conditions, with the NCS2 mounted in the experimental fixture but with no electromagnetic pulse delivered. The image subset used for baseline measurement was identical to the subset used for the corresponding perturbed campaigns, so that any deviation observed in the perturbed runs reflects the effect of the EM pulse rather than dataset-level variation. The measured baseline top-1 and top-5 accuracies on the 512-image validation subset are summarized in Table~\ref{tab:baseline}; these are the values against which the perturbed top-1 accuracies in Sections~\ref{sec:overall-outcomes}--\ref{sec:async} are compared. The baseline measurements were repeated at the start of each experimental day; the inter-day variation of the unperturbed accuracy was below the resolution at which trial-to-trial accuracy is reported in subsequent sections.

\begin{table}[!t]
\caption{Baseline accuracy for each evaluated model on a fixed
512-image subset of the ImageNet validation set (deterministic random
seed), measured under nominal conditions on the Intel NCS2 with the
OpenVINO runtime.}
\label{tab:baseline}
\centering
\begin{tabular}{lccc}
\hline
Model & $N_{\mathrm{img}}$ & $A_{\mathrm{top1}}$ & $A_{\mathrm{top5}}$ \\
\hline
ResNet-18 & 512 & 0.7207 & 0.9004 \\
ResNet-50 & 512 & 0.7813 & 0.9355 \\
VGG-11    & 512 & 0.7090 & 0.8965 \\
\hline
\end{tabular}
\end{table}

\subsection{Overall Fault Outcomes}
\label{sec:overall-outcomes}

The fault-injection campaign reported in this paper comprises the parameter-search trials performed during the exploratory phase and the fixed-parameter trials performed during the spot-test phase, across three models, three probe geometries, and two injection-timing regimes. The exploratory phase contributed several thousand trials, distributed non-uniformly over the parameter space by the TPE sampler, and is not used for the quantitative outcome breakdown reported here because of the sampler-induced bias. The spot-test phase contributed 256 trials per (model, configuration) pair, in which the per-trial inference workload comprised 512 ImageNet validation images.

\begin{figure*}[!t]
  \centering
  \includegraphics[width=0.8\textwidth]{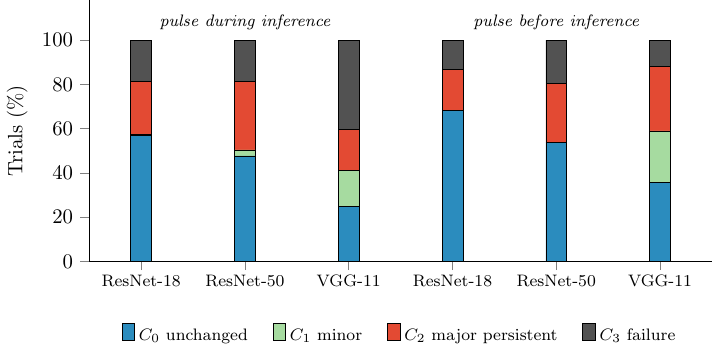}
  \caption{Spot-test fault outcome distribution at the right-side hotspot
  $(x,y,z) = (123.4, 155.1, 0.25)$~mm with the 1\,mm CCW probe at
  $V = 348$~V, 256 trials per (model, timing) pair, 512 inferences per
  trial. The four classes are defined in Section~\ref{sec:fault-classification}: $C_{0}$ unchanged,
  $C_{1}$ minor degradation, $C_{2}$ major persistent degradation, and
  $C_{3}$ device failure. VGG-11 stands out by its large $C_{1}$
  population and high $C_{3}$ rate under during-inference pulses; the
  latter inverts under before-inference pulses.}
  \label{fig:outcomes}
\end{figure*}

The spot-test outcome distribution at the right-side hotspot identified in the exploratory phase, with the 1~mm CCW probe at $(x, y, z) = (123.4, 155.1, 0.25)$~mm and $V = 348$~V, is summarized in Table~\ref{tab:spot-outcomes} and visualized as a stacked bar chart in Fig.~\ref{fig:outcomes} for both injection-timing regimes. Several features of Table~\ref{tab:spot-outcomes} are noteworthy. First, the persistent class~$C_{2}$ is consistently inducible across all three models in both timing regimes, with persistent-fault rates between 18\,\% and 31\,\%. Second, VGG-11 exhibits a class~$C_{1}$ rate that is approximately an order of magnitude higher than either ResNet variant under during-inference pulses, indicating that the architecture is more susceptible to small-magnitude accuracy perturbations that do not collapse the output entirely. Third, the failure rate $R_{\mathrm{fail}}$ for VGG-11 is markedly higher than for the ResNets under during-inference pulses (40\,\% vs.\ 19\,\%), but the ranking inverts under before-inference pulses, where VGG-11 has the lowest failure rate (12\,\%) of any (model, timing) pair in the table. We do not have a confirmed mechanism for the timing-dependent inversion and identify it as a target for follow-up work.

\begin{table}[!t]
\caption{Spot-test outcome distribution at the right-side hotspot
$(x,y,z) = (123.4, 155.1, 0.25)$~mm with the 1~mm CCW probe at
$V = 348$~V, pulse width 160~ns. 256 trials per (model, timing) pair;
each trial comprises 512 inferences. ``During'' indicates a pulse
delivered 1~s after inference start; ``Before'' indicates a pulse
delivered 1~s before inference start. Class definitions are given in
Section~\ref{sec:fault-classification}.}
\label{tab:spot-outcomes}
\centering
\begin{tabular}{llrrrr}
\hline
Model & Timing & $C_{0}$ & $C_{1}$ & $C_{2}$ & $C_{3}$ \\
\hline
ResNet-18 & During &  146 &   1 &  61 &  48 \\
ResNet-50 & During &  122 &   7 &  79 &  48 \\
VGG-11    & During &   64 &  41 &  48 & 103 \\
\hline
ResNet-18 & Before &  175 &   0 &  47 &  34 \\
ResNet-50 & Before &  138 &   0 &  68 &  50 \\
VGG-11    & Before &   91 &  60 &  75 &  30 \\
\hline
\end{tabular}
\end{table}

\subsection{Transient Effects During Inference}
\label{sec:spot-results}

\begin{figure}[!t]
  \centering
  \includegraphics[width=0.48\textwidth]{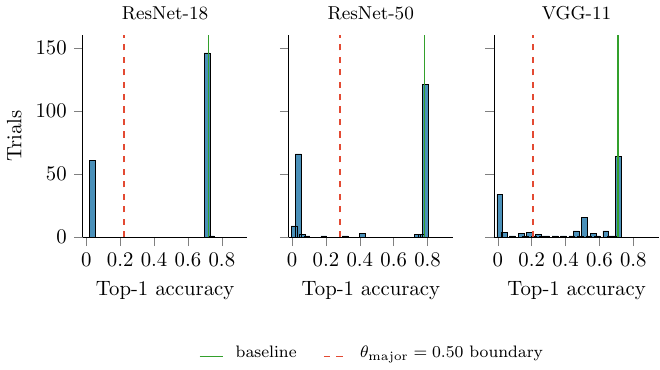}
  \caption{Per-trial top-1 accuracy distributions over 256 spot-test trials
  with one pulse delivered approximately 1~s after inference start at
  $(x,y,z) = (123.4, 155.1, 0.25)$~mm, $V = 348$~V. Class~$C_{3}$ (device
  failure) trials are excluded as they do not produce a valid accuracy.
  The vertical green line in each panel marks the corresponding model's
  unperturbed baseline; the dashed red line marks the threshold
  $\theta_{\mathrm{major}} = 0.50$ separating $C_{1}$ from $C_{2}$. The two ResNet distributions are sharply bimodal,
  populating only the $C_{0}$ and $C_{2}$ regions; the VGG-11 distribution
  spreads substantial probability mass between the two modes, populating
  the $C_{1}$ region as well.}
  \label{fig:bimodal}
\end{figure}

The during-inference spot tests produced a mixture of $C_{0}$, $C_{1}$, $C_{2}$, and $C_{3}$ outcomes. Examination of the per-trial top-1 accuracy distributions, shown in Fig.~\ref{fig:bimodal}, reveals a clearly bimodal structure for the two ResNet models: trials cluster sharply around either the unperturbed baseline accuracy or a near-zero accuracy floor, with the intermediate $C_{1}$ class accounting for at most 3\,\% of trials (1/256 for ResNet-18 and 7/256 for ResNet-50, see Table~\ref{tab:spot-outcomes}). The bimodality is consistent with the interpretation introduced in Section~\ref{sec:fault-classification}: the two modes correspond to the unaffected class~$C_{0}$ and the major-degradation class~$C_{2}$, while the intermediate trials are the rare $C_{1}$ events in which the disturbance flipped a small enough fraction of the output to leave most predictions correct. The VGG-11 distribution, in contrast, exhibits substantially more mass at intermediate accuracies (41/256 trials in $C_{1}$, $\sim 16$\,\%); this is examined in Section~\ref{sec:arch-results}.

The trials assigned to class~$C_{1}$ exhibit two distinct sub-types when the per-image outputs are inspected. In the first sub-type, the deviation from baseline is small in magnitude---typically corresponding to a change in the predicted label for one or two images out of the 512 in the workload---and persists across subsequent inferences within the trial without further degradation; this sub-type is consistent with corruption of a single weight or instruction in a region of the model with high redundancy, where the corruption affects only a small fraction of the input distribution. In the second sub-type, the deviation is confined to the perturbed inference and the next inference on the same image returns the unperturbed prediction; this is the SET-like sub-type proper, and is consistent with corruption of an in-flight activation tensor or an internal scratchpad. Distinguishing these two sub-types operationally requires comparing the perturbed inference output with the next inference on the same image, which the campaign protocol records and which the analysis in this section relies on.

The class~$C_{2}$ trials are unambiguous: top-1 accuracy collapses to near zero on the perturbed inference and remains at near zero across all subsequent inferences within the trial, until the model is explicitly reloaded. The post-collapse top-1 accuracy is concentrated tightly between 0.02 and 0.03 for ResNet-50, between 0.02 and 0.05 for ResNet-18, and between 0.01 and 0.04 for VGG-11; these residual values are above the uniform-random 1000-class chance level and therefore suggest that some corrupted outputs retain residual structure or class bias rather than becoming purely random. The persistence-until-reload signature is the defining diagnostic of class~$C_{2}$ and motivates the SEU-like interpretation in Section~\ref{sec:fault-classification}. The fact that explicit model reload restores baseline accuracy in all observed cases provides further evidence that the persistent corruption resides in loaded model state on the device, rather than in non-volatile storage or in irreversible damage to the silicon.

\subsection{Persistent Effects and SEU-Like Behavior}
\label{sec:persistent-effects}

Class~$C_{2}$ is reproducible across all three evaluated models and across both 1~mm probe orientations, though with different per-trial rates. In every observed instance, an explicit model reload over USB restored baseline accuracy on the next inference; a device-level USB power-cycle was not required, distinguishing $C_{2}$ operationally from $C_{3}$.

\begin{figure*}[!t]
  \centering
  \includegraphics[width=0.8\textwidth]{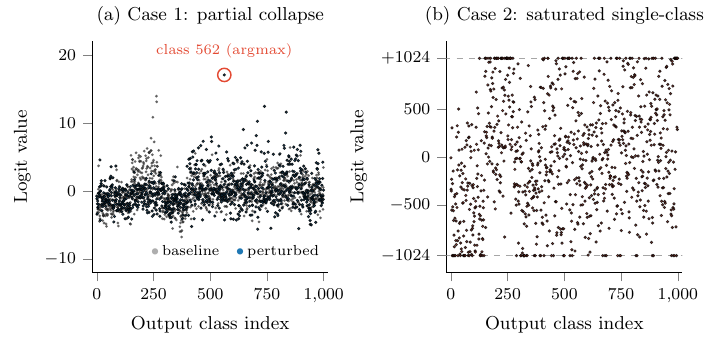}
  \caption{Output-layer logit vectors (1000 ImageNet classes, ResNet-50)
  for representative single inferences in each $C_{2}$ sub-regime.
  \emph{(a) Partial collapse:} perturbed logits remain in the nominal
  numerical range ($\sim [-10, +20]$); the highlighted point at class~562
  is the trial-dominant predicted class, with magnitude only modestly
  above background. \emph{(b) Saturated single-class:} perturbed logits
  span the observed on-device fixed-point range, with 41~classes
  pinned near the positive ceiling and 34~classes pinned near the negative ceiling for this inference. Note the difference in $y$-axis scale between panels: (a)~spans tens, (b)~spans thousands.}
  \label{fig:logits}
\end{figure*}

A closer examination of the raw 1000-element output layer across $C_{2}$ events reveals two distinct sub-regimes that the trial-level top-1 metric does not separate. In the first sub-regime, illustrated in Fig.~\ref{fig:logits}(a), the post-pulse logits remain in their nominal numerical range (approximately $[-10, +30]$) but the argmax distribution across inputs collapses from approximately 640 distinct classes for the unperturbed model to approximately 350, with the single most-frequent class predicted on roughly 8\,\% of inputs (compared with 0.4\,\% under the unperturbed model). Top-1 agreement with the reloaded-model prediction on a per-image basis is approximately 23\,\%, well above the uniform-random 1000-class chance level ($\sim 0.1\,\%$) but well below the 100\,\% agreement expected when comparing the unperturbed model with itself on the same images. This sub-regime is consistent with corruption that biases the classifier toward a small subset of classes without numerically saturating any internal representation, and is responsible for the residual non-zero post-collapse top-1 accuracy reported earlier. In the second sub-regime, illustrated in Fig.~\ref{fig:logits}(b), the post-pulse logits reach the numerical saturation ceiling of the on-device fixed-point representation ($\pm 1023.5$ at the layer reading observed here), and the argmax for every input collapses onto a single class for the entire post-pulse inference sequence. Top-1 agreement with the reloaded-model prediction in this sub-regime is at chance. Across the 22 $C_{2}$ events for which we captured the full post-pulse output trace (ResNet-50 at the right-side hotspot), 10 events showed the partial-collapse sub-regime and 12 showed the saturated-single-class sub-regime; the dominant class in the saturated sub-regime varied between events despite the pulse parameters being held fixed across trials.

The location of the corruption that produces a class~$C_{2}$ outcome cannot be determined from the present data. The persistence-until-reload signature is consistent with at least three hypotheses: corruption of stored model weights in on-die CMX or off-die DDR memory, corruption of the compiled SHAVE-instruction stream loaded with the weights, or corruption of accelerator-internal configuration state (scaling factors, buffer descriptors, DMA control words) set up at model initialization and not refreshed between inferences. The two sub-regimes constrain this hypothesis space: saturation to the $\pm 1023.5$ ceiling is consistent with corruption upstream of, or within, a value path that propagates to the observed output and reaches the numerical dynamic-range limit, while the partial-collapse sub-regime is consistent with corruption of classifier weights, late-stage feature representations, or configuration state whose perturbation biases but does not dominate the output. The across-event variability of the dominant class in the saturated sub-regime, and the distinction between numerically-saturating and numerically-bounded $C_{2}$ events, are both new observations not reported by~\cite{bhasin2025practical} or by prior simulation studies. Discriminating between the three hypotheses would require instrumenting the OpenVINO runtime to dump intermediate tensors after a $C_{2}$ event; this is identified as a planned follow-up.

\subsection{Impact of Injection Timing}
\label{sec:image-count}

The during-inference and before-inference timings produce qualitatively similar outcomes at the right-side hotspot, with class~$C_{2}$ rates of comparable magnitude in the two regimes (Table~\ref{tab:spot-outcomes}). That a pulse delivered before the start of inference can produce class~$C_{2}$ outcomes at appreciable rates is reliability-relevant on its own: it demonstrates that the loaded model state on the NCS2 is fault-susceptible while the device is otherwise idle, and that an integrity check performed only at model-load time is in principle insufficient. The mitigation implication is discussed in Section~\ref{sec:reliability-implications}.

The two timing regimes nevertheless differ in two operational respects. Class~$C_{1}$ events under before-inference pulses are entirely absent for the ResNet models (0/256 trials for both ResNet-18 and ResNet-50), while VGG-11 exhibits an even higher class~$C_{1}$ rate under before-inference pulses than under during-inference pulses (60/256 vs.~41/256). The interpretation we offer for the ResNet observation is that the perturbation pathway available to a before-inference pulse is necessarily through stored state---weights, instructions, or configuration---rather than in-flight activations, and that for the ResNet architectures, persistent corruption of stored state tends either to leave inference unaffected (most weights are sufficiently redundant) or to break it severely (the affected weight or instruction lies on the critical path of most predictions), with little intermediate territory. The class~$C_{3}$ rate is also lower under before-inference pulses for VGG-11 (12\,\% vs.~40\,\%); one possible explanation is that less in-flight control or scheduling state is active at the moment of pulse delivery, although the present data do not directly verify this mechanism.

A separate sensitivity to the inference workload size emerged from a controlled comparison in which the same pulse parameters were applied to ResNet-50 with different numbers of images per trial: 64, 128, 256, and 512. At small workloads (64 and 128 images), the threshold $\theta_{\mathrm{minor}} = 0.01$ used in Section~\ref{sec:fault-classification} falls below the resolution of a single misclassified image---one misclassification in a 64-image workload changes top-1 accuracy by 0.0156, which exceeds $\theta_{\mathrm{minor}}$---and consequently the class~$C_{0}$ count drops to zero by classification artifact rather than by physical change. The class~$C_{2}$ and $C_{3}$ rates are stable across workload sizes within the precision of 64 trials per workload, indicating that the underlying physical fault rate is not a strong function of workload duration over the range examined. We therefore restrict quantitative comparisons elsewhere in this paper to workloads of 512 images, for which the $\theta_{\mathrm{minor}}$ threshold is well-resolved.

A randomized $(x, y, z)$ surface scan with the pulse delivered before inference identified far fewer sensitive locations than the corresponding during-inference scan: in 2750 random points at constant voltage, only three trials produced a clear class~$C_{2}$ outcome. This indicates that the spatial sensitivity of the device to before-inference pulses is more localized than its sensitivity to during-inference pulses, even though at the right-side hotspot the two timings produce comparable rates.

\subsection{Model-Dependent Robustness}
\label{sec:arch-results}

\begin{figure}[!t]
  \centering
  \includegraphics[width=0.48\textwidth]{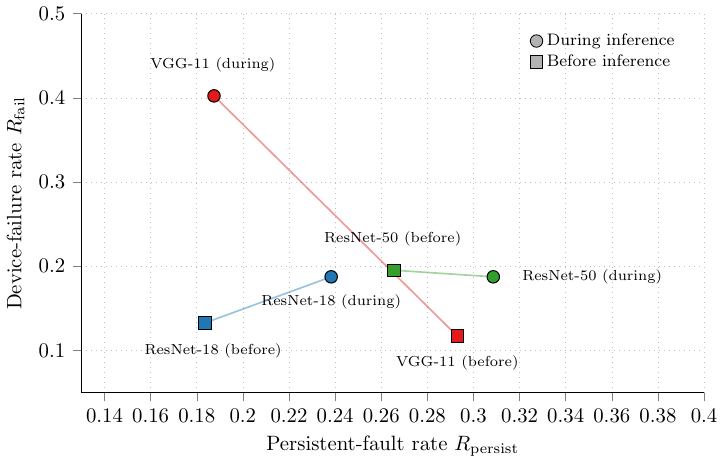}
  \caption{Persistent-fault rate $R_{\mathrm{persist}}$ versus
  device-failure rate $R_{\mathrm{fail}}$ for the six (model, timing)
  configurations of Table~\ref{tab:spot-outcomes}. Each model occupies a
  distinct region of the reliability plane, and the during-vs-before-
  inference pulse timing moves each model along a different vector,
  indicating that timing sensitivity is itself architecture-dependent.}
  \label{fig:persistfail}
\end{figure}

The 256-trial spot tests of Table~\ref{tab:spot-outcomes} support a direct comparison between architectures under identical pulse parameters; the resulting $(R_{\mathrm{persist}}, R_{\mathrm{fail}})$ scatter is shown in Fig.~\ref{fig:persistfail}. Under during-inference pulses, the three models occupy three distinct points in the plane: ResNet-18 at $(0.24, 0.19)$, ResNet-50 at $(0.31, 0.19)$, and VGG-11 at $(0.19, 0.40)$. The two ResNets share the same failure rate but differ in persistent-fault rate, with the deeper ResNet-50 more susceptible to $C_{2}$ outcomes by 7~percentage points. VGG-11 occupies a markedly different region of the plane: a lower persistent-fault rate but a substantially higher failure rate, and a class~$C_{1}$ rate that is, as already noted, approximately an order of magnitude higher than either ResNet.

We offer the following hypothesis, distinguishing it clearly from the measurements above. The lower $C_{2}$ rate and higher $C_{1}$ rate of VGG-11 are jointly consistent with the larger capacity of its fully-connected classifier head: a corruption of a single pre-classifier feature or weight is more easily absorbed by a wide FC layer than by the relatively narrow classifier of a ResNet, so that a fraction of the input distribution continues to be classified correctly even after persistent corruption, producing a $C_{1}$ outcome rather than a $C_{2}$ outcome. The higher $C_{3}$ rate of VGG-11 under during-inference pulses is harder to explain at the architecture level alone and may be associated with its larger memory footprint, longer data movement, or different scheduling behavior, any of which could increase the opportunity for a pulse to disrupt control or communication state. Importantly, neither component of the interpretation can be verified directly from the present measurements, because the campaign cannot localize the corruption to a specific layer; per-layer fault localization, which requires hardware-triggered EMFI, is identified as a methodological prerequisite.

\subsection{Spatial Sensitivity Analysis}
\label{sec:spatial}

\begin{figure*}[!t]
  \centering
  \includegraphics[width=0.8\textwidth]{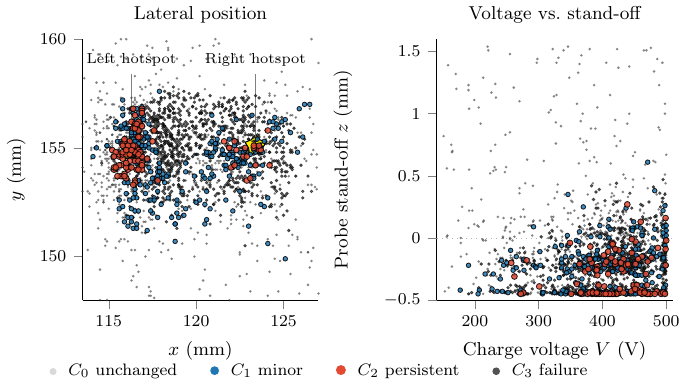}
  \caption{Spatial sensitivity of the NCS2 to a 1\,mm CCW EM pulse during
  ResNet-50 inference, from approximately 16{,}000 Optuna-driven
  exploratory trials. \emph{Left:} Lateral $(x,y)$ map showing two
  reproducible hotspots in which class~$C_{1}$ and class~$C_{2}$
  outcomes are concentrated; the central region between the hotspots is
  dominated by class~$C_{3}$ device failures. \emph{Right:} Voltage--
  stand-off ($V, z$) scatter; classes $C_{1}$ and $C_{2}$ require small
  stand-off ($z \lesssim 0.5$~mm after the probe-specific calibration of
  Table~\ref{tab:probe-zoffsets}) and high voltage ($V \gtrsim 350$~V),
  while class~$C_{3}$ events extend further into the parameter space.
  Markers are layered with $C_{0}$ (light) underneath and $C_{2}$ (red)
  on top to ensure rare interesting events are visible against the dense
  background.
  }
  \label{fig:spatial}
\end{figure*}

The lateral $(x, y)$ sensitivity of the package was mapped using the 1~mm CCW probe with the Optuna-driven exploratory protocol described in Section~\ref{sec:campaign}, with ResNet-50 as the workload and the pulse delivered approximately 1~s after the start of inference. The resulting trial-outcome map is shown in the left panel of Fig.~\ref{fig:spatial}: two reproducible hotspots are visible, in which trials of class~$C_{1}$ and class~$C_{2}$ are concentrated. The hotspots are centered at approximately $(x, y) = (116.3, 154.9)$~mm (the left-side hotspot) and $(x, y) = (123.4, 155.1)$~mm (the right-side hotspot), with a lateral separation of approximately 7~mm. The intermediate region between the two hotspots is dominated by trials of class~$C_{3}$, indicating that pulses delivered to the central region of the package are more likely to disrupt device functionality than to produce a localized data corruption.

This spatial structure is consistent with a die in which functional sub-blocks responsible for control flow and host communication are located centrally while sub-blocks responsible for compute and on-die memory are located laterally: a pulse delivered laterally would then disturb compute or storage state and produce class~$C_{1}$ or $C_{2}$ outcomes, while a pulse delivered centrally would disturb control flow and produce class~$C_{3}$ outcomes. We note this as a hypothesis only; the Myriad~X die floorplan is not publicly available, and a definitive spatial-to-block correspondence would require independent localization (e.g.\ X-ray imaging) which the present study does not perform. The spatial separation of class~$C_{1}/C_{2}$ from class~$C_{3}$ outcomes is, however, directly observed and does not depend on the floorplan interpretation.

\subsection{Voltage and Probe Effects}
\label{sec:repeatability}

The voltage--stand-off ($V, z$) sensitivity, shown in the right panel of Fig.~\ref{fig:spatial}, indicates that successful induction of class~$C_{1}$ and class~$C_{2}$ outcomes is concentrated at small stand-off (below approximately 0.5~mm) and high voltage (above approximately 350~V), with rare events extending to lower voltages. Trial outcomes near the boundary of this region are sensitive to small variations in stand-off and voltage, which contributes to trial-to-trial variability at parameter points near the fault-onset boundary.

\begin{figure}[!t]
  \centering
  \includegraphics[width=0.45\textwidth]{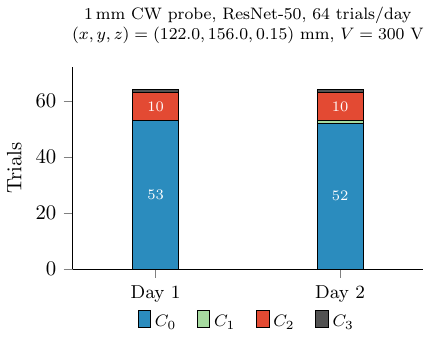}
  \caption{Day-to-day repeatability of the four-class outcome distribution
  at a fixed parameter point with the 1\,mm CW probe, ResNet-50 workload,
  64 trials per session. The two days yield outcome counts that agree to
  within one trial across all four classes, indicating that the
  fault-injection campaign produces stable population statistics at
  parameter points well inside the fault-onset region.}
  \label{fig:repeatability}
\end{figure}

The two 1~mm probes, in CW and CCW orientation respectively, produced qualitatively similar fault distributions when applied to comparable hotspots at matched voltage. A repeatability check on a CW-probe spot-test configuration ($(x,y,z) = (122.0, 156.0, 0.15)$~mm, $V = 300$~V, ResNet-50, 64 trials per session) was performed on two separate experimental days; the resulting outcome distributions, shown in Fig.~\ref{fig:repeatability}, agreed to within one trial across all four classes (Day~1: $C_{0}=53, C_{1}=0, C_{2}=10, C_{3}=1$; Day~2: $C_{0}=52, C_{1}=1, C_{2}=10, C_{3}=1$), indicating that the day-to-day variability anecdotally noted earlier in the campaign was specific to parameter points near the fault-onset boundary rather than a systematic property of the platform.

The 4~mm CCW probe behaves qualitatively differently. Despite covering a substantial fraction of the package surface, the 4~mm probe produces class~$C_{2}$ outcomes only at low pulse voltage (approximately 150~V); at higher voltages with the 4~mm probe, class~$C_{3}$ outcomes dominate to the exclusion of all other classes. The interpretation is straightforward: the larger probe couples energy into a substantially larger area of the die, so that at high voltage the disturbance is delivered simultaneously to multiple functional sub-blocks. One plausible explanation is that the broader disturbance more often affects functionality required for continued execution or host communication, producing a hang rather than a localized data corruption. The 4~mm probe is therefore not a substitute for the 1~mm probes for the present application, although it provides a useful complementary observation about the spatial extent over which the device tolerates simultaneous disturbance.

\subsection{Asynchronous Inference Results}
\label{sec:async}

The during-inference and before-inference experiments were repeated under the OpenVINO asynchronous inference API, in which multiple inference requests are queued and executed in overlap rather than sequentially. The asynchronous execution mode is operationally significant because it more closely reflects the way a production deployment would dispatch inference requests, and because the overlap between data movement and computation may in principle change the observability of an injected fault. The four outcome classes defined in Section~\ref{sec:fault-classification} remain inducible under asynchronous execution, and the spatial structure of the hotspots identified in Section~\ref{sec:spatial} is preserved, with class $C_{1}/C_{2}$ outcomes concentrated at the two lateral hotspots and class~$C_{3}$ dominant in the central region.

A quantitative comparison at the right-side hotspot, with ResNet-50 as the workload at $V = 350$~V and 128 images per trial (a separate matched-parameter campaign run distinct from Table~\ref{tab:spot-outcomes}), yields outcome distributions that differ qualitatively between sync and async execution. Under during-inference asynchronous execution, the class~$C_{0}$ count is essentially zero (every completed trial produces some deviation from baseline) and the class~$C_{1}$ count is correspondingly inflated; the class~$C_{2}$ rate is lower than under matched sync execution while the class~$C_{3}$ rate is markedly higher ($\sim 48\,\%$ vs.\ $\sim 25\,\%$). Under before-inference asynchronous execution the picture inverts: the class~$C_{3}$ rate falls below the matched sync rate while $C_{1}$ climbs to the great majority of trials. The inflation of $C_{1}$ at the expense of $C_{0}$ is consistent with a single pulse interacting with multiple in-flight inference requests at different stages of completion, but the present data do not isolate this mechanism.

\section{Discussion}\label{sec:discussion}

\subsection{Reliability Implications}
\label{sec:reliability-implications}

The four-class fault outcome taxonomy of Section~\ref{sec:fault-classification} separates the observed behavior of the NCS2 along the dimension that matters for a reliability argument: detectability. Class~$C_{3}$ (device hang or runtime error) is less problematic from a silent-data-corruption perspective because it is detectable by host-side liveness checks; a watchdog timer of the kind already standard in industrial deployments is expected to catch this class. Classes~$C_{1}$ and $C_{2}$, in contrast, constitute the silent-data-corruption regime: inference completes, the OpenVINO runtime returns a numerical result, and the host application has no way to distinguish a perturbed output from an unperturbed one without an out-of-band reference.

The intrinsic novelty of class~$C_{2}$ from a reliability standpoint lies not in the per-trial incorrectness but in its temporal envelope: because the corrupted state persists across all subsequent inferences until model reload, the cost of a single $C_{2}$ event is the integrated cost of every silently-incorrect inference until that reload occurs---in many deployment patterns, the full remaining application lifetime.

The pre-inference results (Section~\ref{sec:image-count}) extend this implication: stored model state is fault-susceptible in the idle-loaded condition, so any reliability argument that relies only on a load-time integrity check is insufficient. The required mitigation is periodic re-verification, treated in Section~\ref{sec:mitigation}.

Third, the architecture-dependent ranking (Fig.~\ref{fig:persistfail}) extends the bit-flip-level architecture-dependence of~\cite{li2017understanding, hong2019terminal, reagen2018ares} from simulation to physical perturbation on real silicon, and makes model choice a deployment-time reliability lever for system integrators working under an SDC budget.

\subsection{Mitigation Strategies}
\label{sec:mitigation}

The four-class taxonomy maps directly to a graded mitigation strategy. Each class admits a specific detection or recovery mechanism, and the choice of which classes a deployed system must detect determines the implementation cost.

\textit{Class~$C_{3}$ detection} is the cheapest and is already standard practice. A host-side watchdog with a timeout slightly longer than the longest expected inference latency catches every device hang, USB disconnection, and OpenVINO runtime error in our campaign. The recovery action is a USB power-cycle followed by model reload; we used the \texttt{uhubctl} utility for this purpose during the campaign and confirmed that it restores the device to a clean state in all observed $C_{3}$ instances. No modification to the inference application is required beyond adding the watchdog.

\textit{Class~$C_{2}$ detection} requires either model integrity verification or output cross-checking. The OpenVINO runtime does not currently expose a primitive for reading back loaded weights from the NCS2 over USB, so a pure integrity-hash approach is not directly available without instrumented firmware. A practical alternative is reference-image cross-checking: a small, fixed test image is classified periodically and the predicted top-1 label is compared against a stored golden output. A class~$C_{2}$ event is detected by a reference-image check only if the corruption changes the prediction or checked output features for the selected reference input; using multiple reference images or comparing logits rather than only the top-1 label can increase detection coverage at a constant per-check cost. The detection latency is bounded by the inter-check interval and the corresponding mitigation is a model reload, which we confirmed restores baseline accuracy in every observed case.

\textit{Class~$C_{1}$ detection} is the most demanding because the per-trial accuracy deviation is small and overlaps with the natural variance of the unperturbed system. Reference-image cross-checking detects only those $C_{1}$ events whose perturbation happens to flip the prediction on the specific reference image, and is therefore a low-recall detector. The classical mitigation pattern for this regime is redundant inference: two independent accelerators can provide disagreement detection, while three or more enable majority voting, at the cost of increased inference latency and hardware. For deployment domains with a hard SDC budget set by ASIL or SIL classification, this redundancy cost is justified; for less-stringent domains, accepting an undetected $C_{1}$ rate of a few percent may be acceptable provided the rate is characterized.

We note that all three detection mechanisms above are application-level and can be implemented without modifying the NCS2 firmware or the OpenVINO runtime. Hardware-level mitigations---electromagnetic shielding of the package, supply-rail filtering, on-die error-correcting memory---are in principle possible but lie outside the deployment-time control of a system integrator using a commercial accelerator and are therefore not pursued further here.

\subsection{Security Implications and Scope}

Although this paper is framed and evaluated as a reliability characterization, the same fault classes that motivate the mitigation strategies above also describe the worst-case behavior of the device under an adversarial perturbation source. A reader concerned with security threat modeling can read the four-class taxonomy as a hardness statement: an adversary with physical access to the package and a low-cost commercial EMFI tool can drive the device into class~$C_{2}$ at rates of 18--31\,\% per single pulse at characterized hotspots, and the resulting silent misclassification persists across many subsequent inferences. The mitigation strategies of the previous subsection apply identically in the adversarial setting, with the additional observation that an adversary will preferentially target the $C_{2}$ regime because it is the regime in which the accelerator returns a successful but incorrect result and is therefore the regime that maximizes the cost-asymmetry of the disturbance. We do not pursue an explicit threat-model analysis here because the contributions of the paper are characterization rather than attack, but the result of that characterization is directly usable for security analysis as well.

\section{Threats to Validity}
\label{sec:threats}

We identify four threats to the validity of the present results, in approximate order of how strongly we expect them to limit the conclusions, together with the steps taken to bound each.

The most consequential threat is the \emph{millisecond-scale jitter of the pulse trigger} relative to inference start (Section~\ref{subsec:emfi_platform}). The trigger is issued in software through the ChipSHOUTER's USB serial interface and the inference is started through a separate TCP message to the Raspberry~Pi, with both paths mediated by Python interpreter scheduling and OS-level USB latency. The jitter precludes per-layer fault localization and is the most likely single source of unexplained variance in our class-distribution counts. We bound it by holding the nominal pulse delay fixed within each campaign, by aggregating 256 trials per spot-test parameter point so that per-trial jitter averages into a stable population statistic, and by checking day-to-day repeatability at fixed parameters (Fig.~\ref{fig:repeatability}). The bimodality of the ResNet outcome distributions (Fig.~\ref{fig:bimodal}) and that repeatability check together indicate that the underlying fault rate is itself stable; the jitter limits what we can resolve about the fault \emph{mechanism}, not the fault \emph{rate}.

The second threat is the \emph{single device sample}. All quantitative results derive from one physical NCS2 unit; inter-unit variability is unknown. Manufacturing process variation, package-level differences, and per-unit thermal characteristics can in principle shift the spatial hotspots, the voltage threshold, or the relative class distribution. We did not have a second NCS2 available within the project timeline, and we do not claim that the absolute coordinates and voltages reported in Section~\ref{sec:results} transfer directly to a second unit. We expect the qualitative four-class taxonomy to be more transferable than the absolute coordinates or voltage thresholds, while the architecture-dependent ranking remains a hypothesis to be tested in follow-up work.

The third threat is the \emph{absence of confound controls}. We did not run a matched no-pulse campaign of equal duration to bound the rate at which class~$C_{2}$-like outcomes might arise from causes unrelated to the EM disturbance, such as USB-host drift or thermal-throttling-induced numerical inconsistency. The day-to-day repeatability check of Fig.~\ref{fig:repeatability} provides indirect evidence that the apparatus itself does not generate spurious class-distribution variation at fixed parameters, and the spatial concentration of $C_{2}$ outcomes at characterized hotspots (Fig.~\ref{fig:spatial}) is difficult to explain by a non-spatial confound, but a direct no-pulse control would provide a stronger bound. The ambient laboratory temperature was not directly logged during the campaign, which limits our ability to retroactively quantify thermal-state contributions to per-day variability; this is identified as a measurement that future replication efforts should add.

The fourth threat is the \emph{non-availability of the Myriad~X die floorplan}. The interpretation in Section~\ref{sec:spatial} of the spatial structure---that lateral hotspots correspond to compute and storage blocks while the central $C_{3}$-dominated region corresponds to control logic---is offered as a hypothesis consistent with the observed phenomenology, not as a confirmed correspondence. A definitive spatial-to-block correspondence would require either Intel-internal die documentation or independent localization through, for example, X-ray imaging of the package, neither of which is available to us. Importantly, the four-class taxonomy itself, the persistence-until-reload signature of $C_{2}$, and the architecture-dependent ranking of the three CNN families are all directly observed and do not depend on the floorplan interpretation.

We do not regard the restriction to a single accelerator, to image-classification workloads, or to three CNN architectures as a threat to validity in the same sense, since the paper is explicitly about characterizing this device and these workload families; generalization is identified as future work in Section~\ref{sec:concl}.

\section{Conclusion}\label{sec:concl}

We presented a systematic single-pulse EMFI characterization of the Intel Neural Compute Stick~2 under three production-scale ImageNet classifiers. Pulses at characterized hotspots produce four reproducible outcome classes mapping onto no-effect, SET/SDC-like, SEU-like persistent corruption, and SEFI-like loss of functionality. The campaign extends the prior NCS2 EMFI feasibility study of Bhasin~\textit{et~al.}~\cite{bhasin2025practical} with two findings absent from that work---the persistent $C_{2}$ regime and the pre-inference fault-induction pathway---and establishes architecture choice as a deployment-time reliability lever. The mitigation strategies of Section~\ref{sec:mitigation} follow from the taxonomy and are all implementable at the application level without modification to the device firmware or the OpenVINO runtime.

Several future directions follow from the methodological constraints identified in Section~\ref{sec:threats}. Replacing the software-issued trigger with a hardware GPIO trigger asserted by an instrumented OpenVINO build would enable per-layer fault localization and allow direct testing of the layer-dependent sensitivity profiles predicted by the simulation literature. Replicating the campaign across multiple NCS2 units would bound the inter-unit variability that the present results cannot characterize. Adding matched no-pulse and far-field pulse controls would tighten the bound on confound contributions. Extending the workload set to object detection, depth estimation, and transformer-based inference would test whether the four-class taxonomy is preserved across qualitatively different computation patterns, and extending to other commercial inference accelerators (other VPUs, edge TPUs, embedded GPUs) would test the generality of the architecture-dependent ranking. Finally, implementing and benchmarking the mitigation strategies of Section~\ref{sec:mitigation} would supply quantitative residual fault rates and complete the reliability argument from disturbance source to deployment-ready system.

\subsection*{Acknowledgment}
This work was supported by the Open Research Fund of The State Key Laboratory of Blockchain and Data Security (Grant number: A2566), Zhejiang University.

\subsection*{Usage of AI}
The authors used Anthropic Claude and OpenAI ChatGPT for language polishing and improvement of the presentation of this paper.
All technical content, experiments, analyses, and conclusions were developed and verified by the authors.

\bibliographystyle{IEEEtran}
\bibliography{bib}

\end{document}